# Survey of remnant seasonal water ice patches at southern polar Mars

Gergacz M.[1,2,3], Kereszturi A.[3,4]

1 ELTE Institute of Physics, H-1117 Budapest, Pázmány Péter 1/A, Hungary

2 Wigner RCP, H-1121 Budapest, Konkoly-Thege Miklós 29-33., Hungary

3 Konkoly Thege Miklos Astronomical Institute, Research Centre for Astronomy and Earth Sciences, H-1121 Budapest, Konkoly-Thege Miklós 15-16, Hungary

4 CSFK, MTA Centre of Excellence, H-1121 Budapest, Konkoly-Thege Miklós 15-17., Hungary

**Highlights**

Ice patches were surveyed on Mars left behind the recession of southern polar cap

148 HiRISE images showed ice patches at 40°-60° S lat. and 140°-200° solar long. range

Ice patches were left behind at the shadowed parts of elevations

Many of them behind rocks, which might receive stronger illumination is spring

Formation of microscopic liquid water and brine should be considered there

**Abstract**

On Mars it is possible that after the recession of the seasonal polar ice cap, small icy patches in shady and/or low thermal conductivity places are left behind. These regions are then illuminated by direct sunlight during the summer, warm up and in an ideal case a liquid phase could emerge. This work is surveying HiRISE images for such ice patches and found 148 images with ice patches on them out of the 730 images that fit the selection criteria of location and season. Their separation of ice from other bright patches, like clouds or lighter shades of layers and rocks was possible by their bluish color and strong connection to local topographic shading. Images with ice patches ranged between 140° and 200° solar longitude in the latitude band between -40° and -60°. The diameter of the ice patches ranges between 1.5-300 meters, and they remain on the surface even after the seasonal polar cap has passed over the area for the duration range of 15-139 Martian days.

Based on the Mars Climate Database (MCD) the surface temperature does not reach the melting point of water, and the model indicated no continuous $CO_2$ and $H_2O$ ice cover at the analyzed areas. The ice patches found could be considered such small ice spots that are below the spatial resolution of the model, and their existence was allowed by local shading geometry (decreasing the effect of direct sunlight). The temperature of these ice patches might be even lower than the model suggests because of their elevated albedo compared to the surrounding barren surface. Targeted temperature estimation for these icy patches should be done in the future as a next step. This survey identified such icy patches, which are located behind elevated topographic features, which might be the ideal target for a hypothetic springtime melting by strong solar illumination. Considering the already available information, the possibility exists that below the bulk melting point, premelting process produced microscopic interfacial water between the ice patches and the regolith below.

# 1. Introduction

Water ice melts at 273 K on Mars, but its sublimation temperature is much lower, around 200 K, so as the temperature rises, the ice sublimates much before it can enter the liquid phase. However, in theory it cannot be ruled out that with a fast enough temperature rise, water ice reaches the melting point (Schorghofer 2020.). The aim of this work is to evaluate whether there are such frost patches that might melt in theory during the spring. As the southern polar ice cap retreats, small patches of ice may remain, and the possibility of the liquid phase appearing in them due to increasing irradiance cannot be excluded. Such ice patches can be studied in images from the Mars Reconnaissance Orbiter (MRO) High Resolution Imaging Science Experiment (HiRISE) and their distribution in space and time can be analyzed. In addition, we have analyzed temperature and surface ice volume data for simulated icy regions using the model of the Mars Climate Database (MCD) (Forget et al. 1999., Millour et al. 2018.).

Because the Martian atmosphere and surface has a low thermal conductivity and inertia (Grott et al. 2021., Kereszturi et al. 2012.), small ice patches may remain on the surface by an asymmetrically and discontinuous receding polar cap (Schmidt et al. 2009), in places where they are shielded from the light, for example by slope angles or shadowing surface features. Eventually these areas may also be exposed to sunlight, and then the ice may warm rapidly - it is not yet known whether or not a liquid phase (Schorghofer 2020, Pál and Kereszturi 2017.) may then appear, which is an important question for chemical transformations and the potential for life (deVera et al. 2014., Marschall et al. 2012., Horváth et al. 2009.), while these patches might help to improve near surface humidity estimation (Pal 2019).

If any liquid phase emerges, it might influence low temperature chemical changes on Mars, especially if supported by subzero temperature microscopic liquid water like proposed for hydrogen-peroxide decomposition (Kereszturi, Gobi 2014.) or for sulfate formation (Gobi, Kereszturi 2019.). Such locations might need focused analysis in the future by orbiters monitoring them, which requires specific information on their location, the time period in which ice is present there and a selection of the best ones among them regarding potential chemical changes.

**1.2 Background information**

Beside the polar caps two location types show surface frost cover with temporal changes: those as seasonal cap remnants left behind during the springtime recession, and low-latitude frost patches far away from the seasonal cap. Topography related asymmetry in the recession of the northern polar cap was observed (Benson and James 2005), especially large asymmetries were found at south around the Mountains of Mitchell and in connection with the development of the co-called "cryptic" region (James et al. 1992, Kieffer 2000, James et al. 2001). Large remnant frost patches were observed to stay here until the southern summer, with sublimation away after Ls 315 (Piqueux, S. 2008.) being mostly spectral features of $CO_2$ ice (Calvin W. M. et al. 2009), occasionally closer to 30 degree to the equator (Carrozzo et al. 2009).

There are several locations in the middle and low latitude region, where the surface temperature is consistent with the occurrence of $CO_2$ ice mainly at low thermal inertia dusty areas (Piqueux et al. 2016), which might be composed of micrometer size $CO_2$ ice crystals and forming an optically thin layer. Patches emerge ephemerally between 33°S and 24°N could be observed on pole facing slopes in southern winter periods

(Schorghofer and Edgett 2005), where TES and THEMIS data indicated the bright patches are made of $CO_2$ ice formed during months of accumulation, where $H_2O$ ice could be also present.

## 2. Methods

In this work, optical survey and model based temperature estimation were done. Such ice patches were searched for, that remained on the surface after the seasonal polar ice cap had receded from the given location in the southern hemisphere of Mars. Optical HiRISE images from the MRO spacecraft were analyzed, using the JMars software (Christensen et al. 2009.) between -40° and -70° latitude and 140° and 200° solar longitude (the latter corresponding to the end of southern winter, spring and early summer). The appearance of the features seen in the images were characterized by eye, based on scientific publications and personal experience gained during the work. The data were categorized into images with and without residual ice patches. Their occurrence and characteristics were statistically analyzed.

HiRISE is the high-resolution camera of the Mars Reconnaissance Orbiter spacecraft, which has been orbiting Mars since 2006. The camera's 0.5 m diameter mirror telescope is the largest ever used around another planet and is capable of capturing images of the surface at a resolution of 0.3 m/pixel from an altitude of 300 km. It captures images at around 14-16 local time in three color bands: 400-600 nm (blue-green, B-G), 550-850 nm (red, R) and 800-1000 nm (near infrared, NIR).

There were several small bright features in the surveyed images, thus the following criteria were applied to select the most probable ice patches among them. In the pictures, bright features that do not cast shadows (e.g. are not bright rocks) and are situated at the shady side of the elevations evaluating high resolution topography (depressions or elevated structures could be inferred only from the brightness of the images), which were considered probable ice patches. Very diffuse edge areas, e.g. fogs were not counted, however based on the experience of the authors, small ice patches could be well identified and separated from other features (see the example images in the Results section). Beside the larger number black-and white (red channel) images, color images (central part of the images) were also considered. In color images light patches with a bluish tint were considered probable ice patches if they were located on a shading side of the landform. However white patches that were faint in black and white images but met the other criteria were also observed, but not counted here. Bright spots that cast shadows in sunny areas that are often yellowish in color were not considered to be ice patches, but presumably prominent bright rocks.

The distance between the crocus line (M. Giuranna et al. 2007) and the icy images in solar longitude were calculated in Python. The second half of the diagram were cut off exactly at Ls = 100. After having the crocus line defined, a fifth-degree polynomial fit were conducted on the data using scipy.optimize, creating a curve that fits the picked-out pixels. The distance of the icy images and the crocus line in solar longitude was then calculated with numpy script.

To have a rough first insight to the characteristic temperature values for the locations/seasons, at areas where a suitable bright spot was visible, the average temperatures were simulated at noon and midnight local time using MCD data and calculated the expected amount of water ice and carbon dioxide ice on the surface. MCD is a meteorological database (Forget et al. 1999.) that derives data from numerical simulations of the General Circulation Model (GCM). The GCM models the evolution of the Martian atmosphere over time using numerical data. The resolution of the models produced by the software is quite low (32 pixels/degree) and the data retrieved at the desired point is linearly interpolated from the database, so that temperature fluctuations caused by small ice patches or shaded areas cannot be detected. Nevertheless, it can be used for general mapping of surface temperature, as it works with an accuracy of ± 10 K in connection with the model's 1/32 degree base line (~2 km) scale – see for more detailed description of the reasons for such rough temperature estimation in the Discussion section. Pure ice starts melting on Mars around 273 K, so this temperature would be ideal for the emergence of liquid water in areas where carbon dioxide ice has already completely sublimated from the surface, but because of the general dryness, these water ice patches might also sublimate before melting.

## 3. Results

In the course of the research we analyzed 730 images that met the search parameters, out of which 148 showed ice patches. Small ice patches of interest were found between 140° and 200° solar longitude in the southern spring and summer between -40° and -60° latitude. The results show that ice patches can be distinguished from other bright features, mainly such as rocks by optical appearance. Direct evidence of the presence of water ice is not yet available as these patches are much smaller than the spatial resolution of the CRISM data.

In the IRB and RGB HiRISE color images, the ice is a lighter "cold tint" than its surroundings, typically a bluish-white patch, while the light rocks are usually yellowish. The ice typically follows the shape of the shading surface, or more precisely its shadow. At undulating topography like sand dunes, craters, various pits and hills there are many ice patches, while on flat terrain there are few, and those are in the sides of flatter depressions or in cracks. The ice is always on the poleward side of the shading landforms. The diameters of the patches in the images range from 5 pixels to 1000 pixels (i.e., 1.5 meters to 300 meters), and while the smaller diameters are typically oval, the larger ones tend to be elongated, following the shape of the shadow producing features.

If icy, the white patch has no sharp boundary analyzing at the highest resolution (i.e., there is not much contrast difference between the patch and the surroundings), the ice patches will not cast shadows as expected (often one can see bright patches that cast small shadows, these may be rocks that can be detected by looking carefully). Most of the relevant images are located between -40° and -50° latitude, with smaller ice between the 145°-180° solar longitude seasonal phase (Figure 3).

## 3.1 Ice patch examples

In the A-F panels of Figure 1 the identification criteria formulated above can be observed nicely, the spots are bluish and usually appear in illuminated locations, following the shape of the shading object - after analyzing a few dozen images we feel their identification is reliable. In panel A, it is clear that the ice patches are always located on the southern side of the shading landforms. The 4-80 m ice patches are only on those sides of the 'rings' that are always shielded from the sun. Sand dunes are visible in panel B, which provide shade from the sun. Ice patches are present in these places, again, on the southern side of the shading landforms. The rocky field in panel C offers plenty of shadowy areas as well, dozens of small ice patches are visible in the southern side of these rocks.

A typical patch is seen in panel D, on the shaded side of craters. Bright patches meeting the criteria are very often found in places like that. In many cases these are the places where the patches remain longest after the pole cap has been receded. Panel E shows a good example of larger, elongated ice patches. It can be observed that the elongated shape of the patch is obtained when the shading form does not cause a large level difference. The arc of the largest patch is about 210 m long, it does not even fit into the magnified detail. In contrast, panel F shows examples of smaller, oval patches. The patches of 1.5-3 m are observed at the base of small shading landforms and on the side of a small crater. In panel G, the light patches cast shadows and have a yellowish color, therefore it's safe to assume that these are rocks and not ice patches. At first glance the bright white spot in the middle on panel H might looks like to be ice. However, there is a small bluish-white patch at the southern side, which classifies it as an ice patch. Therefore, the bright white spot is a shadowing form, directly illuminated by sunlight excluding the possibility of it being an ice patch.

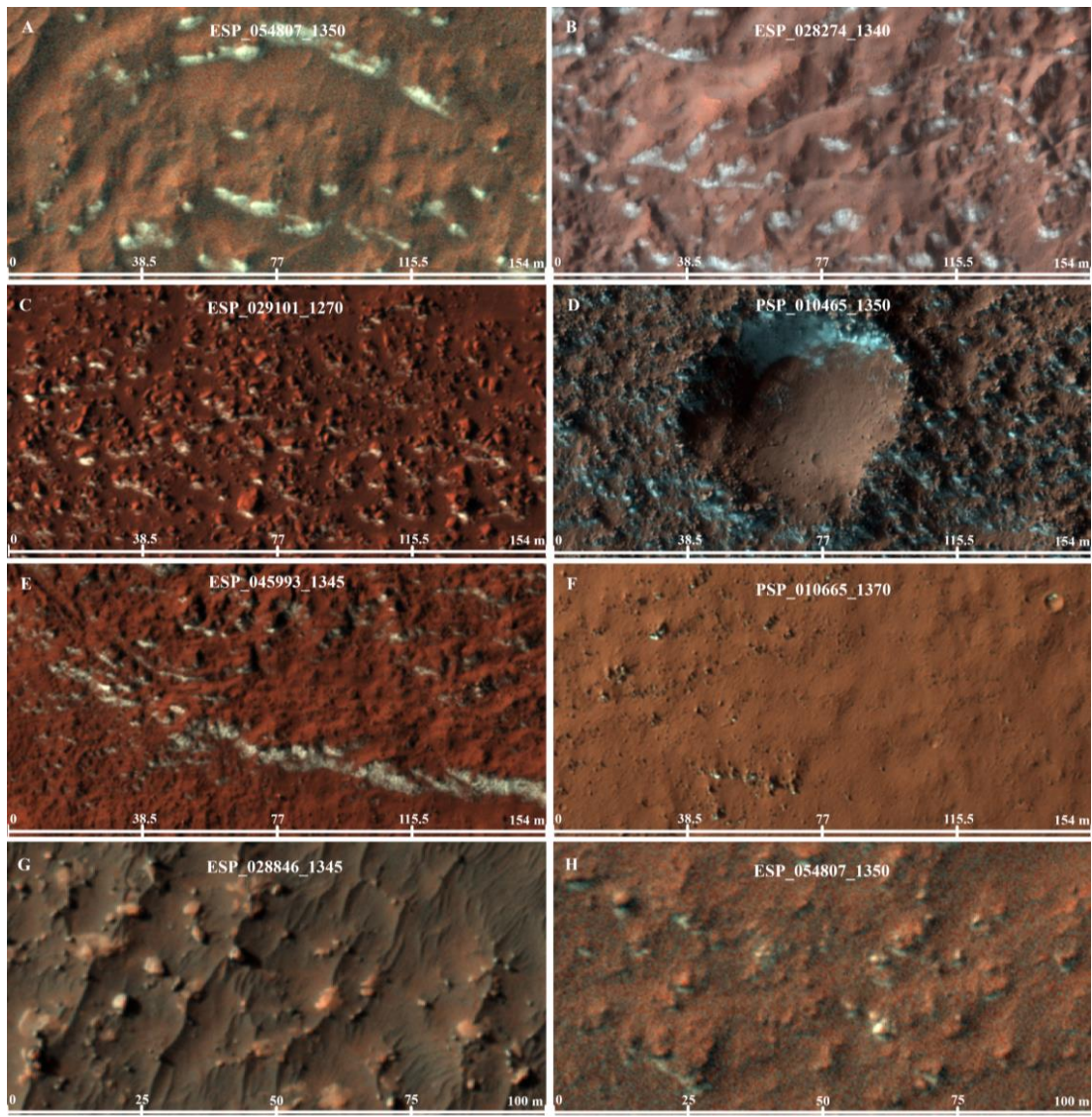

**Figure 1:** Ice patch examples on the southern hemisphere of Mars, with ice patches (excluding panel G where bright rocks are visible), please note that the ice patches occur at the shadowed side of topographic elevations (mainly in C, F, B panels, in C large part of the shadowed inner crater slope is also covered with ice) however there are patches without shadow casting structure also. North is upward in the images.

Further magnified examples for ice patch morphology can be seen in Figure 2 based on ESP_054807_1350 HiRISE image, which shows the following features. Continuous area ice accumulation at the shadow of steep slopes at crater edges (a, d, e, f), at step-like features (c, g, j, k), around stones at small dome-like elevations (b, i), at stones standing separately (n, o, q), and as separated patches around undulating surface elevations (p, r). It is also visible on the images (especially on i, e and f insets) that certain parts of the bight patches are illuminated at the time of image acquisition while other are in shadow.

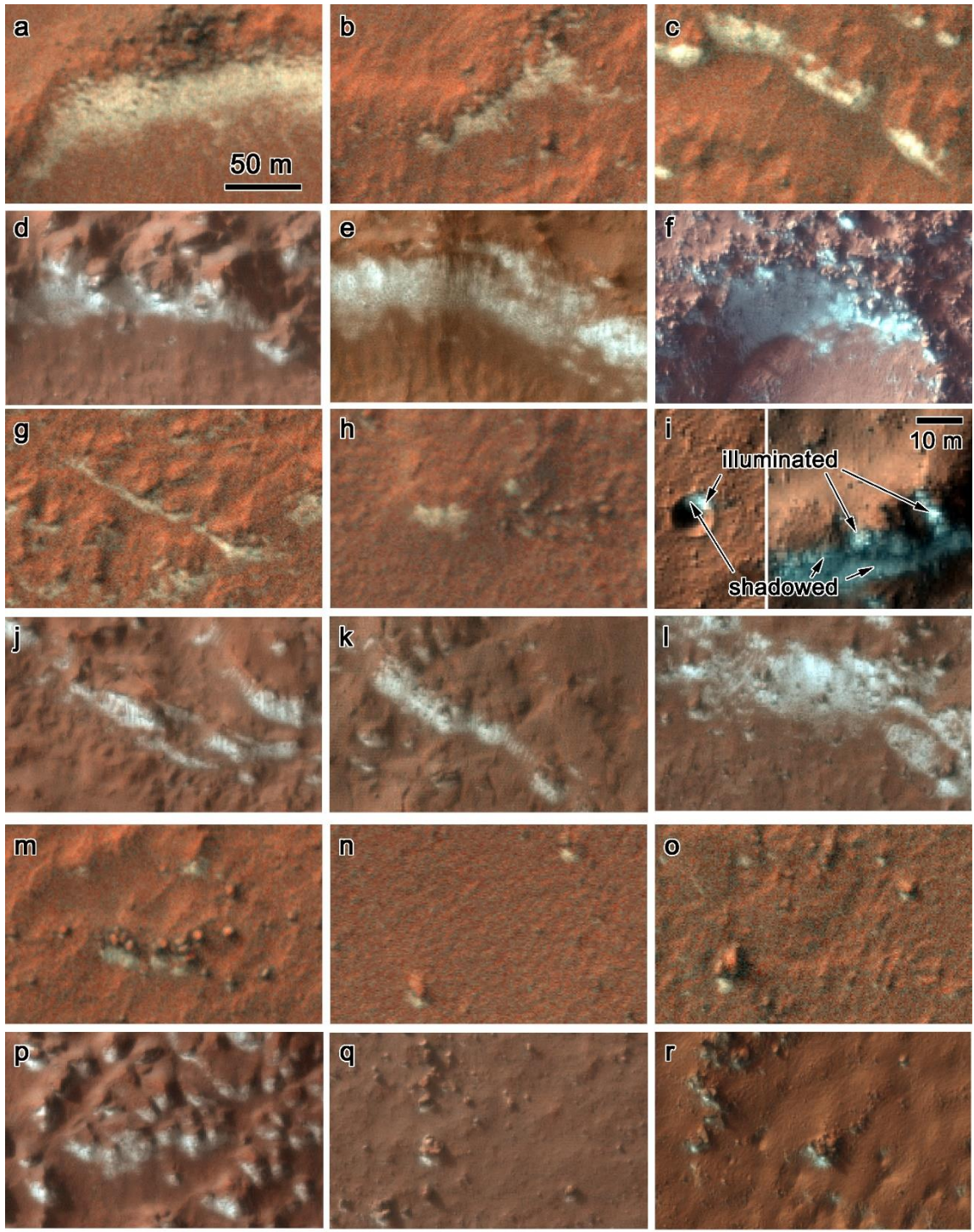

**Figure 2:** Magnified example insets of surface various surface ice patches to visualize their morphology. North s upward in the images.

## 3.2 Summer control image

Summer control tests were also carried out to check the selection criteria. In Figure 3, the right image was taken at Ls = 157.719° in the 32nd Martian year, and shows small patches of ice of 5 to 80 pixels (1.5 to 24 m) on the southern side of the shadowing shapes. The four larger patches (two in 2/G and 2/H) and the other about thirty smaller patches are not present in the left images that were taken at Ls = 260.936°, in 31st Martian year, so we can be confident that the observed ice patches got there in that winter.

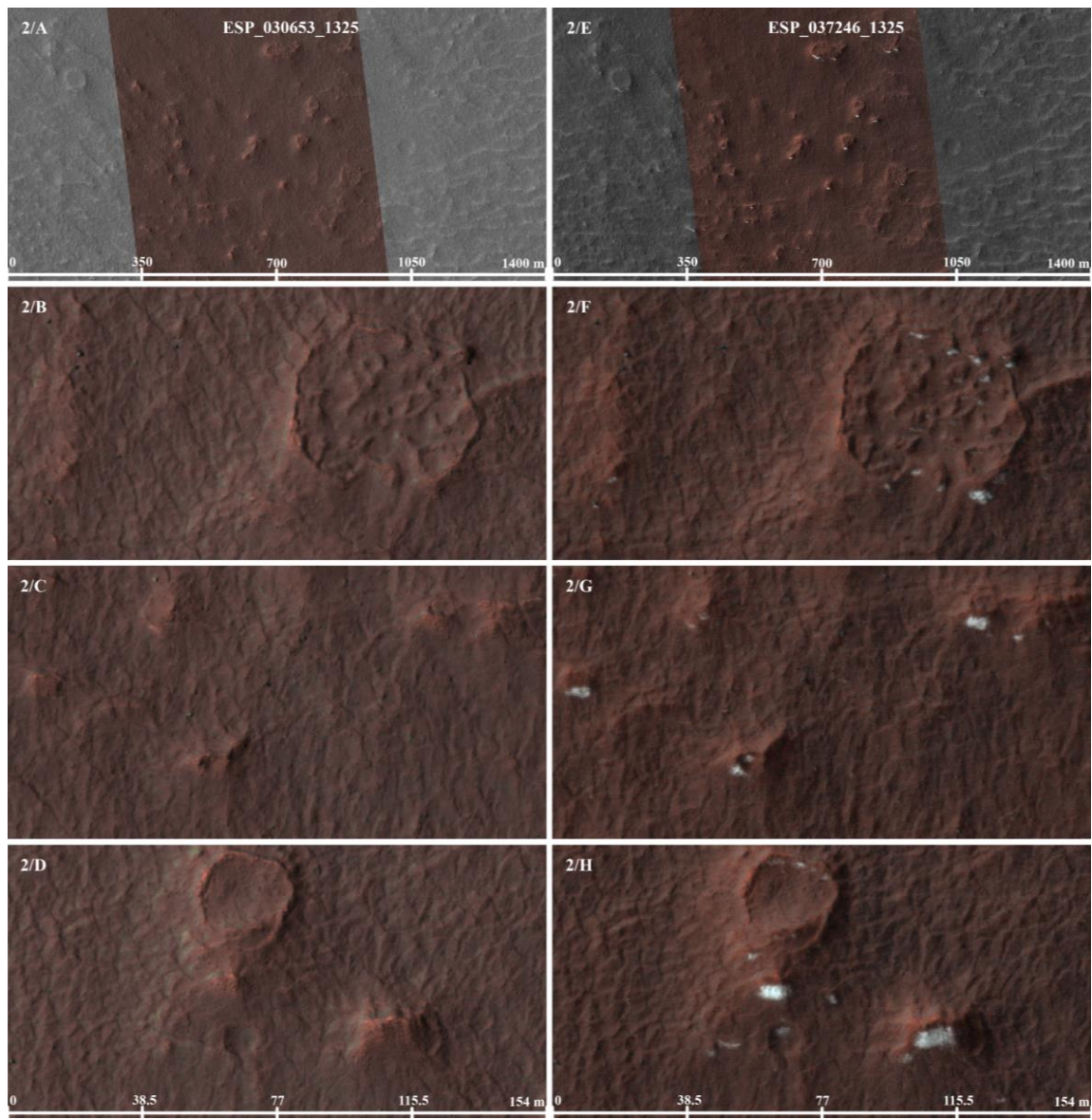

**Figure 3**: The location of ice patches that are not yet present in the summer of the 31st Martian year in the ESP_030653_1325 HiRISE image (2/A) and examples of magnified areas of former ice (2/B, 2/C, 2/D). In the spring of the 32nd Martian year, ice patches appear in the ESP_037246_1325 HiRISE image (2/E) and in the examples of magnified icy areas (2/F, 2/G, 2/H).

**3.3 Location of residual ice patches**

The latitude distribution of images meeting the selection criteria is plotted against the season (solar longitude) the image was taken. Time runs from left to right in the figure, with the edge of the southern polar cap retreating downwards (towards the south pole) accordingly (Frédéric et al. 2009.). Figure 4 shows that with certain lag, the points follow the TES Crocus line, the boundary of the very cold environment where carbon dioxide ice may still exist, measured on 70° longitude (Giuranna et al. 2007.). This suggests that ice patches may remain for a period of about 7.8-72.9 Ls (i.e. 15-139 Martian days) in the regions studied. However, distances in solar longitude were calculated only up to -45.61° latitude, as this is the maximum (closer to the equator, here up to -40°) point of the crocus line. This indicates that some of the detected ice patches are not left behind by the seasonal polar ice cap but were formed in place without building up a continuous ice cap. Such ice patches were observed by others too (Vincendon et al. 2010).

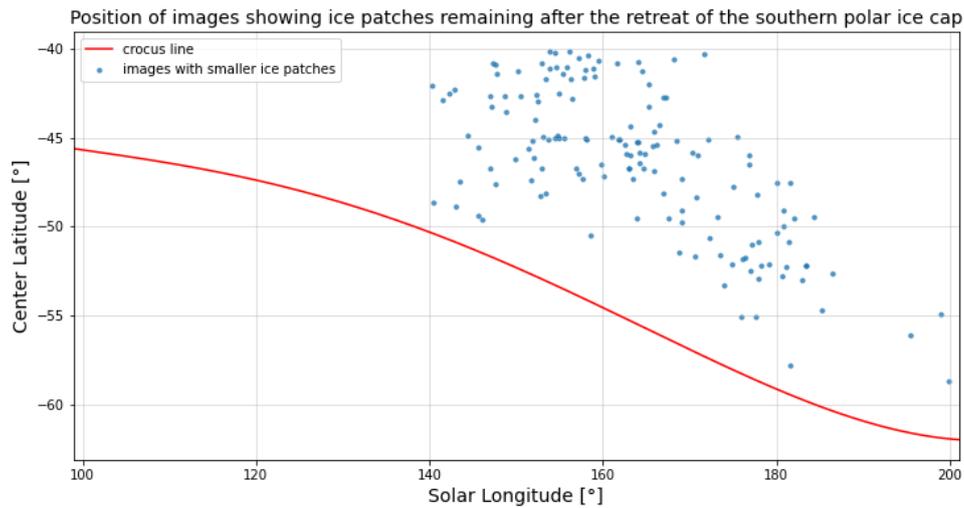

**Figure 4:** Images of ice patches that meet the criteria (blue dots on the graph) follow the TES (Thermal Emission Spectrometer) Crocus line (red line) of the retreating carbon dioxide ice cap, measured on 70° longitude (Giuranna et al. 2007, and from Catalogue).

Comparing images with and without ice patches, the surveyed area needed to be restricted to longitudinal bands instead of global survey (as presented in Figure 5). The survey of positions of icy images were conducted on longitude bands 20° wide where to images without ice patches are also indicated, to prevent overlapping of two types (ice hosting and ice free) the points, since the polar ice cap recedes differently based on geographical longitude.

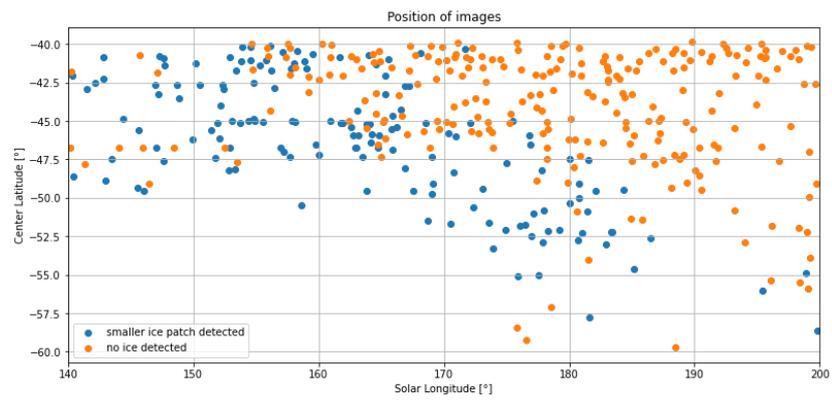
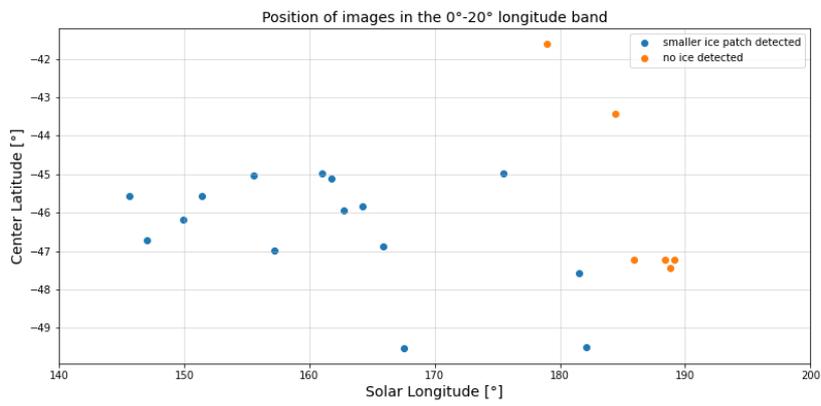
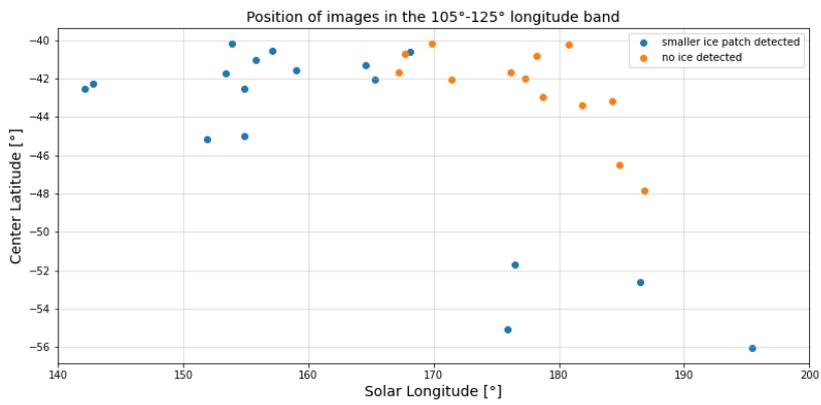
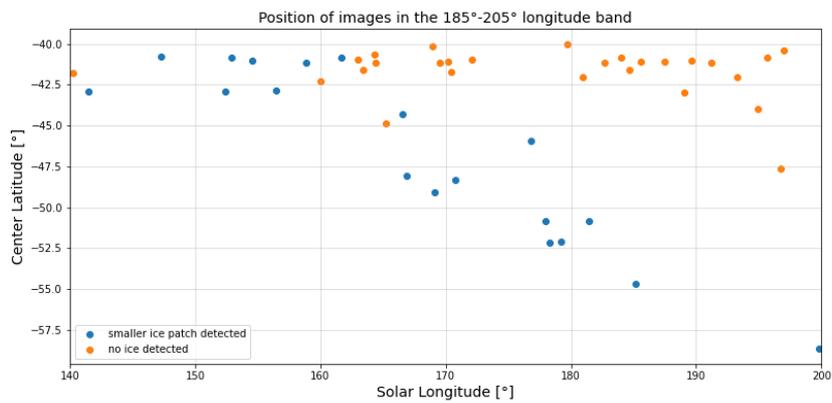

**Figure 5:** Position of images with and without ice patches detected on them on the whole planet and in three different 20° wide longitude bands at the top panel, while the positions of these images are indicated separately for three different latitude bands at the lower three panels. In these three cases the orange and blue dots do not overlap.

The survey at the three restricted longitudinal bands demonstrated that images without ice patches almost never overlap the latitude / Ls range with icy patches, demonstrating that positions of the ice hosting patches according to the location/seasonal period do not overlap with ice patches free images.

3.4 Results of temperature modeling

It is not possible to measure the temperature of certain potential ice patches as they are too small, so we used simulated temperature data from the MCD database, which provides only a very rough approach but still useful information for the targeted area in general. The aim here was to get a general insight to the annual temperature trend at the targeted locations, but for more accurate values, further modelling work is required (please also see the corresponding part of the Discussion section), what goes beyond of the scope of this paper. Here temperatures were simulated for 16 out of the 148 areas with ice patches. On none of the analyzed sites did the modeled temperature reach the ideal surface temperature of 273 K at noon local time, when the HiRISE images were taken. Figure 6 shows an example of the annual temperature curve generated by MCD at latitude -52.6°. The constant temperature of around 140 K is seen during the winter, indicating the presence of the seasonal carbon-dioxide ice polar cap. The model predicts that the retreat starts around Ls = 135°, i.e. at the beginning of the southern spring.

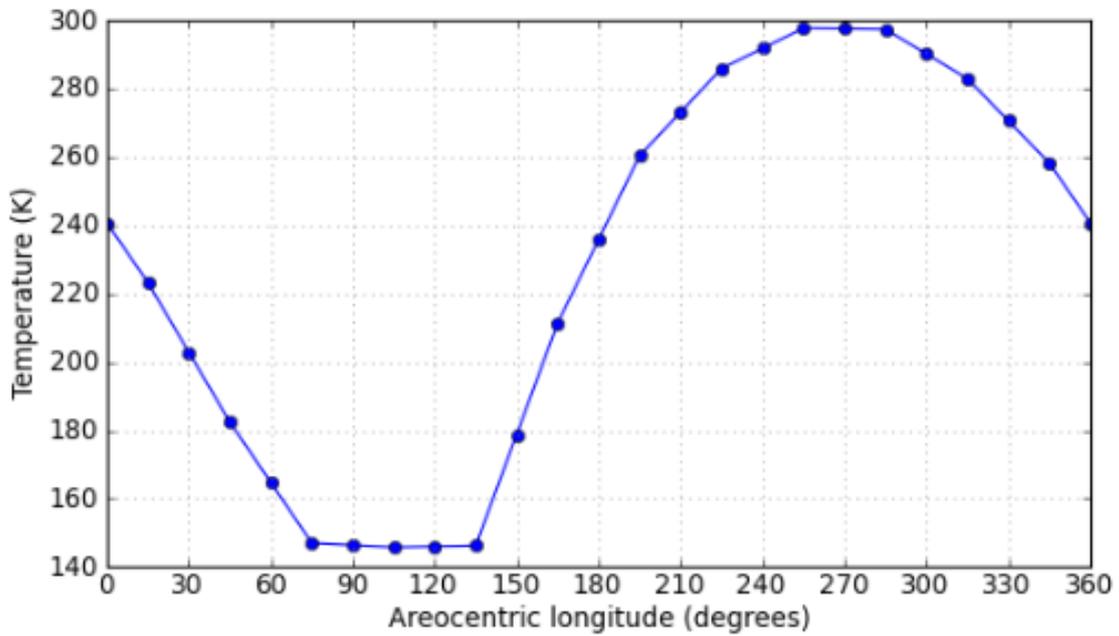

**Figure 6:** Example of a full-year temperature curve at -52.6° latitude. Between 75° and 135° solar longitude, a constant temperature of around 140 K can be seen in winter, which is ideal for $CO_2$ to condense.

The model calculations also suggest that carbon-dioxide ice has already sublimated away in most cases by the time of the study.

**Table 1:** Simulated midnight (T0h) and noon (T12h) temperatures for areas identified as having ice patches and areas difficult to identify. The first column shows the image identifier.

| ID | HiRISE id. number | Latitude [°] | Solar Longitude [°] | $T_{0h}$ [K] | $T_{12h}$ [K] |
|---|---|---|---|---|---|
| 4 | PSP_002862_1195_COLOR | -60,13855 | 195,956 | 158 | 190 |
| 10 | ESP_038210_1210_COLOR | -58,6539 | 199,807 | 172 | 217 |
| 11 | ESP_011338_1270_RED | -52,7331 | 180,68 | 178 | 235 |
| 13 | ESP_011421_1300_RED | -49,4841 | 184,354 | 185 | 242 |
| 15 | ESP_028274_1340_COLOR | -45,5803 | 151,417 | 170 | 218 |
| 16 | ESP_028651_1370_RED | -42,72975 | 166,918 | 185 | 240 |
| 18 | ESP_029101_1270_RED | -52,61755 | 186,477 | 178 | 247 |
| 20 | ESP_037458_1355_COLOR | -44,2742 | 166,52 | 182 | 230 |
| 21 | ESP_045993_1345_RED | -45,00805 | 154,869 | 170 | 224 |
| 24 | ESP_054807_1350_COLOR | -44,89 | 154,804 | 177 | 210 |
| 25 | ESP_055333_1330_COLOR | -46,5023 | 176,892 | 181 | 240 |
| 28 | PSP_010465_1350_RED | -44,8742 | 144,419 | 166 | 209 |
| 31 | PSP_010665_1370_COLOR | -42,61685 | 152,358 | 175 | 226 |
| 32 | PSP_010758_1395_RED | -40,1296 | 156,121 | 180 | 235 |
| 34 | ESP_028742_1315_RED | -48,31965 | 170,778 | 180 | 225 |
| 36 | PSP_010548_1385_RED | -41,3765 | 147,689 | 172 | 230 |
| 38 | ESP_037246_1325_COLOR | -47,3113 | 157,719 | 170 | 212 |
| 49 | ESP_055200_1190_COLOR | -60,8589 | 171,156 | 146 | 148 |

## 4. Discussion

Remnant ice patches after the recession of the seasonal polar cap at the southern hemisphere were identified in the latitude band between -40° and -60°, which could be separated to other bright patches (mainly rocks) based on their appearance and nearby context. Out of the analyzed 730 HiRISE images, 148 showed light patches that matched the criteria. The size of these ice patches ranges between 1.5-300 meters and they follow the shading form. They remain on the surface for about 15-139 Martian days after the seasonal polar ice cap has passed. Several frost patches can be observed at latitude closer to the equator than the maximal extent of the seasonal polar cap, where the frost seems to be also condensed in winter, but the conditions did not allow to cover regionally large (km scale and above) areas. Thus, here the ice condensation never produces a continuous seasonal cap under the current climatic conditions.

Considering the morphology of the ice patches most of them were separated, isolated patches with few m diameter, clearly at the shadowed area of topographic elevations, but there were several cases with continuous (100 m size and above), elongated frost patches at slopes and crater rims. These observations indicate that the main determining factor of ice patches formation is the shadowed location, and the formation of one patch does not help much the enlargement of the given patch. However, they could easily form as larger continuous frost along crater rims or other elongated topographic features. The ice patches surveyed in this work are probably closely related to those patches, which were discovered closer to the equator. For one comparison see Figure 7.

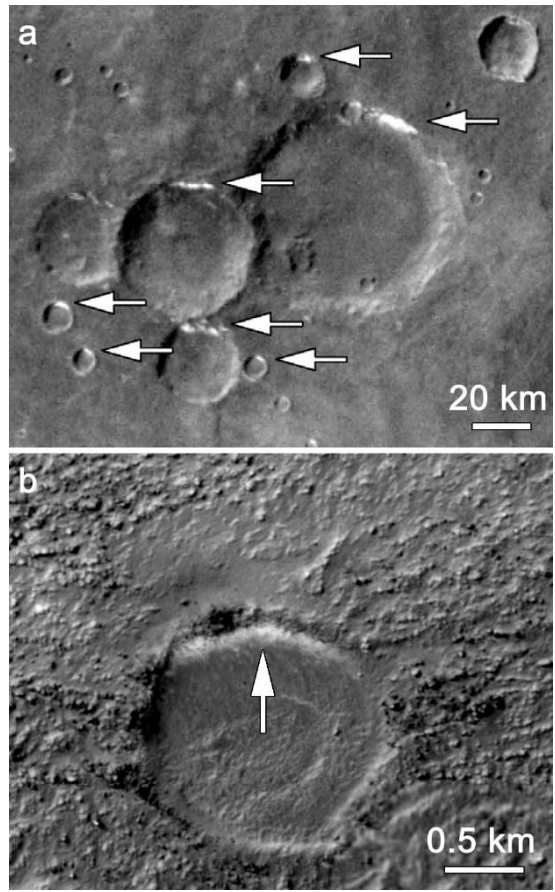

Figure 7. Comparison of MOC (R02-00411) Mars Global Surveyor image (a) with bright frost patches at crater slopes (a), and a HiRISE (ESP_054807_1350) image of a similar but smaller frost patch (b).

The **separation of $CO_2$ and $H_2O$** or even their mixture is not easy as the observed frost patches are below the spatial resolution of THEMIS and even CRISM spectrometers. Considering that there are such illuminated parts of bright frost, which are still present on the surface with 17 degrees solar elevation above the horizon indicates that the ice is there not under grazing solar illumination, indicating these patches might be more elevated temperature than what $CO_2$ frost requires – however this is not a firm evidence. Regarding the temperature aspects, the available modelling method is not enough to gain a realistic estimation for the observed ice patches. In areas with ice patches, the **average surface temperatures** (calculated at the spatial resolution of 32 pixels/degree) were between 175 and 226 K during the day (as seen in Table 1), with the highest simulated temperature being 247 K. Considering subpixel mixing, having a moderately large defrosted and dark area, the small ice patches could be cold enough to be composed of $CO_2$ ice.

The average temperature difference between the melting point of the water ice and the surface temperature at noon local time was thus 47 K on the average in the ice patch containing areas, which still inhibits the melting of pure water ice. Beside this difference, the modelled values are relevant for the defrosted darker surface, while a small but high albedo frost patch could have even lower temperature. Regional differences from tilt and exposure direction of slopes could influence not only stability and occurrence of surface but possible subsurface ice also, especially at high surface roughness locations (Aharonson and Schorghofer 2006). The checked temperature model assumes that the target areas in the considered seasonal periods is ice free, and even if the model provided values taken to be real, there might be subpixel "cold spots". As a result the separation of the two ice type requires further work, as well as the temperature estimation for them using their albedo beside the other parameters.

The temperature values are also might be influenced by the elevated temperature of the nearby (barren) regolith too, as it could warm up the area of ice patches (especially along their outer edge) by heat conduction from the nearby dark and warmed up regolith. The heat conduction inside the otherwise porous regolith might be enhanced by the pore filling absorbed $H_2O$ (Kossacki 2002.), which is expected to be present after the recession of the seasonal $CO_2$ polar cap and seasonal $H_2O$ cap (Kiri et al. 2008.) but before the desiccation of the regolith. Such subsurface adsorbed water ice containing ring was identified in the northern hemisphere (Kuzmin et al. 2009.) but has not been identified in the southern. If it exists, it would increase the heat conductivity in the shallow regolith, and thus the temperature of the regolith at the area of the ice patches. However, for specific evaluation of the temperature there, high resolution and sophisticated modeling work is needed. The best conditions for possible ephemeral melting are present at those ice patches that left behind protruding elevations, which are present based on the observations described in this work. Favorable slope attitude might also contribute in the warming up process.

### 4.1. Possible changes at the identified ice patches

It is worth to mention, whichever ice are the observed patches made of, where $CO_2$ ice is present on Mars, it usually cold raps $H_2O$ ice from the atmosphere. So water ice is highly probable to be present in these patches, and might be left behind for a while after $CO_2$ ice has been sublimated away. Regarding melting possibility, it depends not only on the geographic location but also on the nearby meter scale topographical setting. It could be assumed that those ice patches survive the longest, which keep a shadowed position for the longest period, and receive moderately high illumination when they became exposed to sunlight somewhat "abruptly".

Assuming the ice patches are made of $CO_2$ together with cold-trapped $H_2O$, the later might left behind during the warming (possible in invisible form as $H_2O$ ice might be too thin and transparent but also allow more elevated temperature because of lower albedo). While the melting point of bulk and pure water ice might not be reached, liquid phase as brine or as pure water in the microscopic scale might still emerge there. As a hypothetical approach, using the modeled noon temperatures seen in Table 1 (except ID 4 and 49), these are above the melting point of magnesium perchlorate. The freezing point of the relevant brine solutions in the Martian environment (Chevrier and Altheide 2008) solution and melting point values: $NaClO_4$ 240 K, $CaCl_2$ 223 K, $Mg(ClO_4)_2$ 198 K, $NaClO_4$-$Mg(ClO_4)_2$-$H_2O$ 180. Brines might emerge in the case of sufficient salts in the top regolith layer (Möhlmann 2011, Möhlmann and Thomsen 2011), which might exist all around the planet by the global mixing by winds, what looks to be realistic on current Mars for chlorates (Toner and Catling 2018) and sulfates (Chevrier and Altheide 2008, Chevrier et al. 2022). Microscopic scale is also an important level, where based on Van der Waals forces (Möhlmann 2008.), a thin liquid layer could exist much below the bulk melting point. However, these are only theoretical possibilities, but already show the importance of these icy patches.

Considering further theoretical possibilities, if microscopic liquid water emerges, it might have some limited chemical consequence on the regolith. According to Kereszturi and Gobi (2014) hyperoxide could be decomposed by the emergence of liquid there already at subzero temperatures, at the given location. This process might especially work if the remnant ice patches emerge at the same location in subsequent years. Sulfate formation might also emerge there if the specific components are present (Gobi and Kereszturi 2019.). However, these are usually very low speed reactions, e.g. in reality there might be too few reactant molecules produced to be observable. But under the generally stable Martian surface conditions, if the dust deposition does not influence the areas of these ice patches, the recurring short periods are favorable for the accumulation of small annual chemical effects. All these topics are waiting for further evaluation and in ideal case next missions could have a chance to confirm or reject the related possibilities.

## 5.     Conclusion

After analyzing 730 HiRISE images, ice patches were searched for between -40° and -70° latitude, during the southern spring and summer after the retreat of the seasonal polar ice cap. Altogether 148 images with ice patches on them were observed in the 140° to 200° solar longitude seasonal interval. Such ice patches were almost always observed on the poleward side of shadowing elevations, typically without sharp boundaries. Their identification is enhanced by examining RGB imagery, as the hue of the patch is also important. The size of the observed patches ranges between 1.5-300 meters. Ice can be distinguished from rocks based on their color on the image and whether they seem to cast shadow or not. The observed ice patches nicely follow the TES based Crocus line. On average these patches remain present for 70 days after the receding polar ice cap edge, with the maximum observed time being 139 Martian days. In some cases, such ice patches might form beyond the maximal edge of the seasonal polar cap, indicating these are familiar patches to even lower latitude observed frost patches.

According to temperature models, the average surface temperature at the analyzed areas does not reach 273 K, which is necessary for bulk water ice to melt, and the observed ice patches are likely to sublimate away from the surface completely before that temperature. In all but one of the 16 modeled ice patch areas, the local time noon temperature was 200 K or in one case even 247 K when the HiRISE images were taken. However, the above listed values are only rough modelling based ones for barren surfaces (these MCG based values are relevant for flat areas), the real temperatures might be different and the simulated values are not accurate and relevant enough to draw firm conclusion. Because of their low albedo, these ice patches might be even colder than mentioned above.

A further question is whether the ice patches consist of water ice or carbon dioxide ice. Carbon dioxide ice sublimates at a much lower temperature (about 140 K depending on pressure (Manning et al. 2006.) than water ice (220-240 K on Mars depending on humidity). In the ice patches identified both ice could be present, their separation requires further work, but since these patches are small their separation is not an easy process. However, even if they are composed of $CO_2$, they probably trap cold water vapor from the atmosphere, which might be left behind there with more elevated sublimation temperature. And considering the microscopic scale, melting of water ice is not impossible along the ice-mineral interface due to van der Waals forces, making the identified ice patches an interesting target for future research.

## 6. Acknowledgement

The authors thank the Wigner Scientific Computing Laboratory for their support.